\documentclass{article}

\usepackage{arxiv}

\usepackage[utf8]{inputenc} 
\usepackage[T1]{fontenc}    
\usepackage{hyperref}       
\usepackage{url}            
\usepackage{booktabs}       
\usepackage{amsfonts}       
\usepackage{nicefrac}       
\usepackage{microtype}      
\usepackage{lipsum}

\usepackage{amsmath}
\usepackage{amssymb,graphicx}
\usepackage{color}
\usepackage{verbatim}
\usepackage[dvipsnames]{xcolor}
\usepackage{empheq}
\usepackage{wrapfig}

\title{Light scattering from stationary $\mathcal{PT}$-symmetric collections of particles}

\author{
  Olga Korotkova  \\
  Department of Physics\\
  University of Miami\\
  Coral Gables, FL 33146 \\
  \texttt{korotkova@physics.miami.edu} \\
   \And
 Paulo A. Brand\~ao \\
  Universidade Federal de Alagoas\\
  Instituto de F\'isica\\
  Maceió, Alagoas, 57072-900, Brazil \\
  \texttt{paulo.brandao@fis.ufal.br} \\

}

\begin{document}
\maketitle

\begin{abstract}
A potential scattering theory from deterministic and random $\mathcal{PT}$ collections of particles with gain and loss is introduced and the forms of their structure and pair-structure factors are  elucidated. An example relating to light scattering from a random distribution of a pair of particles with gain and loss is considered.    
\end{abstract}


The problem of weak scattering of electromagnetic fields from conventional particulate collections is fairly well explored \cite{BW}- \cite{Wang2018}. The understanding of principles and mechanisms behind waves - particulate media interaction has led to a number of discoveries, starting from the classic problem of sky color interpretation \cite{Rayleigh}, and up to the recently solved phase problem in crystallography \cite{WolfPRL}. Depending on the deterministic or random nature of the collection and the  number of incident and scattered directions involved in calculation of scattered field properties, it must be characterized by the structure factor \cite{Ziman} or the pair-structure factor \cite{SahinKor2008}. The models of the structure factors for passive (zero gain/loss) collections are well known, e.g., Percus-Yievick model for hard spheres \cite{PY} or the Fischer-Butford model for fractal aggregates \cite{B}.

However, very little is known about the properties of particulate collections with $\mathcal{PT}$ symmetry \cite{bender1998} and those of electromagnetic fields scattered from them. Several recent publications revealed such outcomes for a single electric dipole \cite{Staliunas} and other pairs of gain/loss centers \cite{16}-\cite{18}. The aim of this Letter is to conceptually introduce  deterministic and random  $\mathcal{PT}$-symmetric particulate collections with arbitrary number of centers, obtain the expressions of their structure and pair-structure factors and illustrate the scalar light scattering outcomes from a generic collection of this class. Moreover, to our knowledge, there are no papers treating random $\mathcal{PT}$-symmetric collections. There is in fact a sole paper available in the literature on the continuous stationary $\mathcal{PT}$-symmetric media \cite{BK}. 

For scattering purposes a deterministic realization of a  material occupying volume $V$, at position $\textbf{r}$ and angular frequency $\omega$ can be characterized by the potential \cite{BW}
\begin{equation}
F(\textbf{r},\omega) = (k^2/4\pi^2)[n^2(\textbf{r},\omega) - 1],
\end{equation}
$k$ and $n$ being the radiation's wave number and the particles' refractive index. Under the  first Born approximation, the cross-spectral density $W^s(r\hat{s}_1,r\hat{s}_2,\omega)$ of radiation scattered to the far zone of a stationary medium is given by the integral \cite{Wolf2007} 
\begin{equation}\label{prop1}
\begin{split}
    W^s(r\hat{s}_1,&r\hat{s}_2,\omega) = \frac{1}{r^2}\int_{V}\int_{V} W^i(\textbf{r}_1,\textbf{r}_2,\omega)C(\textbf{r}_1,\textbf{r}_2,\omega) \\
    &\times\exp[-i k(\hat{s}_2\cdot\textbf{r}_2 - \hat{s}_1\cdot\textbf{r}_1)]d^3r_1 d^3r_2
\\&=
\frac{1}{r^2} \widetilde{W}^i(\textbf{r}_1,\textbf{r}_2,\omega)\circledast \widetilde{C}(\textbf{r}_1,\textbf{r}_2,\omega)\Biggr|_{(-k\hat{s}_1,
    k\hat{s}_2,\omega)},
\end{split}
\end{equation}
where $W^i(\textbf{r}_1,\textbf{r}_2,\omega)$ is the cross-spectral density of the incident field, $\circledast$ denotes convolution in six dimensions and
\begin{equation}\label{C}
C(\textbf{r}_1,\textbf{r}_2,\omega)=\langle F^*(\textbf{r}_1,\omega)F(\textbf{r}_2,\omega)\rangle_{pc}
\end{equation}
is the correlation function of the scattering potential while tilde denotes its spatial Fourier transform 
\begin{equation}\label{int}
    \widetilde{C}(\textbf{K}_1,\textbf{K}_2,\omega) = \int_V C(\textbf{r}_1,\textbf{r}_2,\omega)\exp(-i\textbf{K}_1\cdot\textbf{r}_1 - i\textbf{K}_2\cdot\textbf{r}_2)d^3r_1 d^3r_2.
\end{equation}
The average denoted by $\langle \cdot \rangle_{pc}$ in \eqref{C} is taken over the ensemble of collection's realizations. Hence $\widetilde{C}$ plays the crucial role in re-correlating the incident fields (see also \cite{KorWolf2007}). In particular, for an incident polychromatic plane wave with spectral density $S^i(\omega)$, and propagating along direction $\hat{s}_0$, its cross-spectral density is 
\begin{equation}\label{PW}
    W^i(\textbf{r}_1,\textbf{r}_2,\omega) = S^i(\omega)\exp[i k\hat{s}_0\cdot(\textbf{r}_2 - \textbf{r}_1)].
\end{equation}
Substituting from \eqref{PW} into \eqref{prop1} yields 
\begin{equation}\label{Ws}
\begin{split}
    W^s(r\hat{s}_1,r\hat{s}_2,\omega) =S^i(\omega) \widetilde{C}(-\textbf{K}_1,\textbf{K}_2,\omega)/r^2,
    \end{split}
\end{equation}
where $\textbf{K}_l = k(\hat{s}_l - \hat{s}_0)$, $l=1,2$, 
are the momentum transfer vectors characterizing  scattering from incident direction $\hat{s}_0$ to outgoing direction $\hat{s}_l$. In cases when $W^i$ involves more than one direction, the momentum transfer vectors become: $\textbf{K}_1 = k(\hat{s}_1 - {\hat{s}'}_1)$ and $\textbf{K}_2 = k(\hat{s}_2 - {\hat{s}'}_2)$. The spectral density $S^s(\textbf{r},\omega)$ of the scattered field is readily obtained from \eqref{Ws} by setting  $\textbf{r}_1 = \textbf{r}_2 = \textbf{r}$ which implies $\textbf{K}_1 = \textbf{K}_2 = \textbf{K}$. 

We will now attempt to design a single realization of the intrinsically $\mathcal{PT}$-symmetric collection. In this case, the scattering potential $F(\textbf{r},\omega)$ must satisfy the condition \cite{L} 
\begin{equation}\label{FPT}
F^*(\textbf{r},\omega)=F(-\textbf{r},\omega).
\end{equation}
Assume that such a collection consists of $N$ pairs of particles with gain and loss and the collection's center is at point $\textbf{r}=0$. We start by writing the general potential
\begin{equation}\label{dettwo}
        F(\textbf{r},\omega) = \frac{1}{2}\sum_{n=1}^N f(\textbf{r}-\textbf{r}_n^G,\omega) + \frac{1}{2}\sum_{n=1}^N f^*(\textbf{r}-\textbf{r}_n^L,\omega),
\end{equation}
where $f(\textbf{r},\omega)$ is any  complex-valued function representing the scattering potential for each particle and $\textbf{r}_n^{G,L}$ are centres of particles with gain and loss. In order for \eqref{dettwo} to represent a $\mathcal{PT}$-symmetric distribution, \eqref{FPT} must be satisfied. On directly applying condition  (\ref{FPT}) in \eqref{dettwo} one sees that this could be accomplished in two ways: either 
\begin{equation}
f^*(\textbf{r},\omega)=f(-\textbf{r},\omega), \quad \textbf{r}_n=0, \quad \text{for all} \quad  n,
\end{equation}
i.e., $f(\textbf{r},\omega)$ must be $\mathcal{PT}$ symmetric and all particles must be centered at origin $\textbf{r}=0$; or 
\begin{equation}\label{condpt}
f(-\textbf{r},\omega) = f(\textbf{r},\omega), \quad \textbf{r}_n^G = -\textbf{r}_n^L = \textbf{r}_n, \quad \text{for all} \quad  n, 
\end{equation}
i.e., $f$ must be even (in the sense that both real and imaginary parts are even) and the particles with gain and loss must be symmetrically centered around the origin. It immediately implies that the former choice represents a single particle at $\textbf{r}=0$ and hence must be discarded. For the latter choice the three-dimensional Fourier transform of $F(\textbf{r},\omega)$ becomes
\begin{equation}\label{tildeF}
\begin{split}
    \widetilde{F}(\textbf{K},\omega) &= \int F(\textbf{r},\omega)\exp(-i\textbf{K}\cdot\textbf{r})d^3r \\
    &= (1/2)[\widetilde{f}(\textbf{K},\omega) + \widetilde{f}^*(-\textbf{K},\omega)]\sum_{n=1}^N\cos(\textbf{K}\cdot\textbf{r}_n)\\
    &+(i/2)[\widetilde{f}^*(-\textbf{K},\omega) - \widetilde{f}(\textbf{K},\omega)]\sum_{n=1}^N\sin(\textbf{K}\cdot\textbf{r}_n),
\end{split}
\end{equation}
where $\widetilde{f}(\textbf{K},\omega)$ is that of $f(\textbf{r},\omega)$. Since the Fourier transform of an even function is even, we get
\begin{equation}\label{tildeF2}
\begin{split}
    \widetilde{F}(\textbf{K},\omega) &= \Re[\widetilde{f}(\textbf{K},\omega)]\sum_{n=1}^N\cos(\textbf{K}\cdot\textbf{r}_n) \\& -\Im[\widetilde{f}(\textbf{K},\omega)]\sum_{n=1}^N\sin(\textbf{K}\cdot\textbf{r}_n), 
\end{split}
\end{equation}
being a real-valued function. On substituting from \eqref{tildeF2} into Eqs. (\ref{int}) and (\ref{Ws}) [\eqref{C} and \eqref{prop1} without the averages], we obtain the means of predicting the statistics of light scattered from deterministic $\mathcal{PT}$-symmetric collections. 
Passive collections are conventionally characterized by their \textit{structure factors} $\mathfrak{S}(\textbf{K},\omega)$ \cite{Ziman}. For a  $\mathcal{PT}$-symmetric collection 
\begin{equation}\label{SF}
\begin{split}
    \mathfrak{S}&(\textbf{K},\omega)=[\widetilde{F}(\textbf{K},\omega)]^2 \bigl/ |\widetilde{f}(\textbf{K},\omega)|^2\\
    &=\mathfrak{s}_{R}(\textbf{K},\omega)\mathfrak{S}_{R}(\textbf{K},\omega)+\mathfrak{s}_{I}(\textbf{K},\omega)\mathfrak{S}_{I}(\textbf{K},\omega) \\
    &-\mathfrak{s}_{RI}(\textbf{K},\omega)\mathfrak{S}_{RI}(\textbf{K},\omega), 
    \end{split}
\end{equation}
where the trio of \textit{partial structure factors} 
\begin{equation}
\begin{split}
&\mathfrak{S}_{R}(\textbf{K},\omega)=\biggl[\sum_{n=1}^N\cos(\textbf{K}\cdot\textbf{r}_n)\biggr]^2,  \mathfrak{S}_{I}(\textbf{K},\omega)=\biggl[\sum_{n=1}^N\sin(\textbf{K}\cdot\textbf{r}_n)\biggr]^2,\\&
\mathfrak{S}_{RI}(\textbf{K},\omega)=\sum_{n=1}^N\cos(\textbf{K}\cdot\textbf{r}_n) \sum_{n=1}^N\sin(\textbf{K}\cdot\textbf{r}_n)
\end{split}
\end{equation}
is introduced to characterize the distribution of centers and  
\begin{equation}
\begin{split}
&\mathfrak{s}_{R}(\textbf{K},\omega)=\frac{\Re^2[\widetilde{f}(\textbf{K},\omega)]}{|\widetilde{f}(\textbf{K},\omega)|^2}, \quad 
\mathfrak{s}_{I}(\textbf{K},\omega)=\frac{\Im^2[\widetilde{f}(\textbf{K},\omega)]}{|\widetilde{f}(\textbf{K},\omega)|^2}, \\&
\mathfrak{s}_{IR}(\textbf{K},\omega)=2\frac{\Re[\widetilde{f}(\textbf{K},\omega)]\Im[\widetilde{f}(\textbf{K},\omega)]}{|\widetilde{f}(\textbf{K},\omega)|^2}
\end{split}
\end{equation}
are coefficients relating to the properties of individual particles.  

Further, the \textit{pair-structure factor} $\mathfrak{P}(\textbf{K}_1,\textbf{K}_2,\omega)$ of the $\mathcal{PT}$-symmetric collection can be defined as \cite{SahinKor2008}, \cite{SahinKor2009}
\begin{equation}\label{PSF}
\begin{split}
    \mathfrak{P}(\textbf{K}_1,\textbf{K}_2,\omega)&=\widetilde{F}^*(\textbf{K}_1,\omega)\widetilde{F}(\textbf{K}_2,\omega) \bigl/ |\widetilde{f}(\textbf{K}_1,\omega)||\widetilde{f}(\textbf{K}_2,\omega)| \\& =\mathfrak{p}_{R}(\textbf{K}_1,\textbf{K}_2,\omega)
   \mathfrak{P}_{R}(\textbf{K}_1,\textbf{K}_2,\omega) \\&
    +\mathfrak{p}_{I}(\textbf{K}_1,\textbf{K}_2,\omega)
    \mathfrak{P}_{I}(\textbf{K}_1,\textbf{K}_2,\omega)\\& -\mathfrak{p}_{RI}(\textbf{K}_1,\textbf{K}_2,\omega)\mathfrak{P}_{RI}(\textbf{K}_1,\textbf{K}_2,\omega) \\& -\mathfrak{p}_{IR}(\textbf{K}_1,\textbf{K}_2,\omega)\mathfrak{P}_{IR}(\textbf{K}_1,\textbf{K}_2,\omega)
    \end{split}
\end{equation}
Here, the quad of \textit{partial pair-structure factors} is
\begin{equation}\label{PPP}
\begin{split}
&\mathfrak{P}_{RR}(\textbf{K}_1,\textbf{K}_2,\omega)=\sum_{n=1}^N\cos(\textbf{K}_1\cdot\textbf{r}_n)\sum_{n=1}^N\cos(\textbf{K}_2\cdot\textbf{r}_n), \\& \mathfrak{P}_{II}(\textbf{K}_1,\textbf{K}_2,\omega)=\sum_{n=1}^N\sin(\textbf{K}_1\cdot\textbf{r}_n)\sum_{n=1}^N\sin(\textbf{K}_2\cdot\textbf{r}_n),\\&
\mathfrak{P}_{RI}(\textbf{K}_1,\textbf{K}_2,\omega)=\sum_{n=1}^N\cos(\textbf{K}_1\cdot\textbf{r}_n)\sum_{n=1}^N\sin(\textbf{K}_2\cdot\textbf{r}_n), \\&
\mathfrak{P}_{IR}(\textbf{K}_1,\textbf{K}_2,\omega)=\sum_{n=1}^N\sin(\textbf{K}_1\cdot\textbf{r}_n)\sum_{n=1}^N\cos(\textbf{K}_2\cdot\textbf{r}_n)
\end{split}
\end{equation}
with coefficients 
\begin{equation}\label{pp}
\begin{split}
&\mathfrak{p}_{R}(\textbf{K}_1,\textbf{K}_2,\omega)= \Re[\widetilde{f}(\textbf{K}_1,\omega)]\Re[\widetilde{f}(\textbf{K}_2,\omega)]/|\widetilde{f}(\textbf{K}_1,\omega)||\widetilde{f}(\textbf{K}_2,\omega)| \\&
\mathfrak{p}_{I}(\textbf{K}_1,\textbf{K}_2,\omega)=\Im[\widetilde{f}(\textbf{K}_1,\omega)]\Im[\widetilde{f}(\textbf{K}_2,\omega)]/|\widetilde{f}(\textbf{K}_1,\omega)||\widetilde{f}(\textbf{K}_2,\omega)| \\&
\mathfrak{p}_{RI}(\textbf{K}_1,\textbf{K}_2,\omega)=\Re[\widetilde{f}(\textbf{K}_1,\omega)]\Im[\widetilde{f}(\textbf{K}_2,\omega)]/|\widetilde{f}(\textbf{K}_1,\omega)||\widetilde{f}(\textbf{K}_2,\omega)| \\&
\mathfrak{p}_{IR}(\textbf{K}_1,\textbf{K}_2,\omega)=\Im[\widetilde{f}(\textbf{K}_1,\omega)]\Re[\widetilde{f}(\textbf{K}_2,\omega)]/|\widetilde{f}(\textbf{K}_1,\omega)||\widetilde{f}(\textbf{K}_2,\omega)|
\end{split}
\end{equation}

Let us now turn to the concept of a random $\mathcal{PT}$-symmetric collection. Recall that, in general, random collections of identical passive particles could be of four types: (I) centers are deterministic, individual potentials are random; (II) centers are random, individual potentials are deterministic; (III) both centers and potentials are random, two processes are uncorrelated; (IV) both centers and potentials are random, two processes are correlated (the most general type). This classification also holds for particles with gain and loss. In this case, in addition, one must investigate under what conditions the individual realizations formed by particles with gain and loss, whether or not these individual realizations having $\mathcal{PT}$-symmetry, form random collections with or without $\mathcal{PT}$-symmetry. In the most rigorous case we must assume that the joint probability function of any order or the correlation functions of any order for the entire collection are $\mathcal{PT}$-symmetric.  However, for practical purposes we will restrict the attention to stationary collections for which it suffices to make assumptions regarding the first two statistical moments. Indeed we will refer to a collection as stationary, $\mathcal{PT}$-symmetric, if both the average value 
$A(\textbf{r},\omega)=\langle F(\textbf{r},\omega)\rangle_{pc}$ 
and the second-order correlation function of the scattering potential $C(\textbf{r}_1,\textbf{r}_2,\omega)$ in \eqref{C} satisfy the conditions 
\begin{equation} \label{Fav}
A(\textbf{r},\omega)=A^*(-\textbf{r},\omega),
\end{equation}
and \cite{BK}
\begin{equation}\label{CorPT}
C(\textbf{r}_1,\textbf{r}_2,\omega)=C^*(-\textbf{r}_1,-\textbf{r}_2,\omega).
\end{equation}
 
The relation between the $\mathcal{PT}$ symmetry of individual realizations and that of the averaged quantities is not straightforward. Generally, the following holds: (A) If all realizations are $\mathcal{PT}$-symmetric, then $A$ and $C$ must be such as well, and can be degenerated to passive in the limiting case; (B) if at least some realizations are not $\mathcal{PT}$-symmetric, any outcome is possible. 

In case (A), the realizations of the scattering potential constituting the ensemble must be of the form given in \eqref{dettwo} satisfying condition (\ref{condpt}).
Then it follows from \eqref{tildeF2} that the Fourier transform of the correlation function $C$ is given by 
\begin{equation}
\begin{split}
\widetilde{C}(\textbf{K}_1,\textbf{K}_2,\omega)&=\langle \widetilde{F}^*(\textbf{K}_1,\omega)\widetilde{F}(\textbf{K}_2,\omega) \rangle_{pc} \\&    
= \langle |\widetilde{f}(\textbf{K}_1,\omega)||\widetilde{f}(\textbf{K}_2,\omega)| 
\mathfrak{P}(\textbf{K}_1,\textbf{K}_2,\omega) \rangle_{pc}.
\end{split}
\end{equation}
In particular, with the assumption that individual potentials are deterministic, and using \eqref{PSF}, we find that 
\begin{equation}
    \begin{split}
\widetilde{C}(\textbf{K}_1,\textbf{K}_2,\omega)&=  |\widetilde{f}(\textbf{K}_1,\omega)||\widetilde{f}(\textbf{K}_2,\omega)| \\& \times \Bigl[ \mathfrak{p}_{R}(\textbf{K}_1,\textbf{K}_2,\omega)
   \langle \mathfrak{P}_{R}(\textbf{K}_1,\textbf{K}_2,\omega)\rangle_c \\&
    +\mathfrak{p}_{I}(\textbf{K}_1,\textbf{K}_2,\omega)
    \langle \mathfrak{P}_{I}(\textbf{K}_1,\textbf{K}_2,\omega) \rangle_c\\& - \mathfrak{p}_{RI}(\textbf{K}_1,\textbf{K}_2,\omega) \langle \mathfrak{P}_{RI}(\textbf{K}_1,\textbf{K}_2,\omega) \rangle_c \Bigr],
\end{split}
\end{equation}
where the subscript $c$ denotes average over particles' locations.

We will now consider the more general case (B) in which the individual realizations can be either $\mathcal{PT}$-symmetric or not but $A$,$C$ must be such. In order to do so, suppose that a collection consists of $N$ particles with gain and $N$ particles with loss that can take any positions in a three-dimensional space. The individual realizations are still of the form given in \eqref{dettwo} but, unlike in case (A), no assumptions are imposed on $f$'s or $\textbf{r}_n$'s. Then  
\begin{equation}\label{Cgen}
\begin{split}
C(\textbf{R}_1,\textbf{R}_2,\omega)&=\frac{1}{4}\sum_{n,m=1}^N\Biggl\{  \Big< f^*(\textbf{R}_1-\textbf{r}^G_n,\omega)f(\textbf{R}_2-\textbf{r}_m^G,\omega)  \Big>_{pc}
\\& + \Big< f(\textbf{R}_1-\textbf{r}_n^L,\omega)f^*(\textbf{R}_2-\textbf{r}_m^L,\omega )  \Big>_{pc}\\&
+ \Big< f^*(\textbf{R}_1-\textbf{r}_n^G,\omega )f^*(\textbf{R}_2-\textbf{r}_m^L,\omega )\Big>_{pc} \\& + \Big< f(\textbf{R}_1-\textbf{r}_n^L,\omega )f(\textbf{R}_2-\textbf{r}_m^G,\omega )\Big>_{pc} \Biggr\}.
\end{split}
\end{equation}
It was demonstrated in Ref. \cite{BK} that the Fourier transform for any 
$\mathcal{PT}$-symmetric correlation function $C(\textbf{R}_1,\textbf{R}_2,\omega)$ [defined as in \eqref{int}] satisfies the condition:
\begin{equation}\label{PTcond} \widetilde{C}_{PT}(\textbf{K}_1,\textbf{K}_2,\omega) = \widetilde{C}_{PT}(-\textbf{K}_1,-\textbf{K}_2,\omega).
\end{equation}
Substitution from \eqref{Cgen} into \eqref{int} yields 
\begin{equation}\label{Ck1k2}
\begin{split}
    \widetilde{C}&(\textbf{K}_1,\textbf{K}_2,\omega)   \\& =\frac{1}{4}\sum_{nm}\Bigg[ \Big< \widetilde{f}^*(-\textbf{K}_1,\omega)\widetilde{f}( \textbf{K}_2,\omega)\exp(-i\pmb{K}_1\cdot
    \textbf{r}_n^G - i\textbf{K}_2\cdot\textbf{r}_m^G)\Big>_{pc}\\&  +\Big< \widetilde{f}^*(-\pmb{K}_1,\omega)\widetilde{f}^*(-
    \textbf{K}_2,\omega)\exp(-i\textbf{K}_1\cdot \textbf{r}_n^G - i\textbf{K}_2\cdot\textbf{r}_m^L)\Big>_{pc} \\& +\Big< \widetilde{f}(\textbf{K}_1,\omega)\widetilde{f}(\textbf{K}_2,\omega)\exp(-i\textbf{K}_1\cdot\textbf{r}_n^L - i\textbf{K}_2\cdot\textbf{r}_m^G)\Big>_{pc} \\& +\Big< \widetilde{f}(\textbf{K}_1,\omega)\widetilde{f}^*(- \textbf{K}_2,\omega)\exp(-i\textbf{K}_1\cdot \textbf{r}_n^L - i\textbf{K}_2\cdot\textbf{r}_m^L)\Big>_{pc} \Bigg],
\end{split}
\end{equation}
implying that the condition expressed by \eqref{PTcond} is not satisfied in general. To insure it, certain sufficient conditions may be implemented, e.g.,
\begin{equation}
    \textbf{r}_n^G =  -\textbf{r}_n^L = \textbf{r}_n \quad \text{and} \quad    \widetilde{f}(\textbf{K},\omega) \in \mathbb{R}.
\end{equation}
 The second of these conditions is automatically satisfied if $f(\textbf{r},\omega)$ has $\mathcal{PT}$ symmetry. Then, if we assume that $f(\textbf{r},\omega)$ is $\mathcal{PT}$-symmetric, it is also true that $\widetilde{f}(\textbf{K},\omega) = \widetilde{f}^*(-\textbf{K},\omega) = \widetilde{f}(-\textbf{K},\omega)$. Under these conditions, \eqref{Ck1k2} yields
\begin{equation}\label{CPT}
\begin{split}
    \widetilde{C}&_{PT}(\textbf{K}_1 ,\textbf{K}_2,\omega) = \frac{1}{2} \Bigg<\widetilde{f}(\textbf{K}_1,\omega)\widetilde{f}(\textbf{K}_2,\omega) \\& \times \sum_{n,m=1}^N \Bigg[  \cos(\textbf{K}_1\cdot\textbf{r}_n + \textbf{K}_2\cdot\textbf{r}_m)  + \cos(\textbf{K}_1\cdot\textbf{r}_n - \textbf{K}_2\cdot\textbf{r}_m) \Bigg]\Bigg>_{pc} \\&= \Bigg<\widetilde{f}(\textbf{K}_1,\omega)\widetilde{f}(\textbf{K}_2,\omega)\sum_{n,m=1}^N \cos(\textbf{K}_1\cdot\textbf{r}_n)\cos(\textbf{K}_2\cdot\textbf{r}_m)\Bigg>_{pc}.
\end{split}
\end{equation}
If the correlation properties of individual particles are independent from their centers' statistics then $\widetilde{C}_{PT}$ in \eqref{CPT} factorizes: 
\begin{equation}\label{type3}
\widetilde{C}(\textbf{K}_1,\textbf{K}_2,\omega)
=  \widetilde{c}(\textbf{K}_1,\textbf{K}_2,\omega) \mathfrak{P}(\textbf{K}_1,\textbf{K}_2,\omega)
\end{equation}
where 
\begin{equation}\label{tildecc}
 \widetilde{c}(\textbf{K}_1,\textbf{K}_2,\omega)=  \langle \widetilde{f}(\textbf{K}_1,\omega)\widetilde{f}(\textbf{K}_2,\omega) \rangle_p
\end{equation}
and
\begin{equation}\label{psfrandom}
    \mathfrak{P}(\textbf{K}_1,\textbf{K}_2,\omega)= \sum\limits_{n=1}^{N}\sum\limits_{m=1}^{N}\Bigl< \cos(\textbf{r}_n \cdot \textbf{K}_1)\cos(\textbf{r}_m \cdot \textbf{K}_2) \Bigr>_c.
\end{equation}
Here, subscripts $c$ and $p$ signify that two different averages are obtained. When $\textbf{K}_1=\textbf{K}_2=\textbf{K}$, 
\begin{equation}
 \widetilde{c}(\textbf{K},\textbf{K},\omega)=  \langle |\widetilde{f}(\textbf{K},\omega)|^2 \rangle_p,  
\end{equation}
and
\begin{equation}\label{SFrandom}
\mathfrak{S}(\textbf{K},\omega) = \sum\limits_{n=1}^{N}\sum\limits_{m=1}^{N}\Bigl< \cos(\textbf{r}_n \cdot \textbf{K}) \cos(\textbf{r}_m \cdot \textbf{K}) \Bigr>_c.
\end{equation}
For deterministic  $\widetilde{f}(\textbf{K},\omega)$ the averages $\langle \cdot \rangle_p$ are dropped.

\begin{figure}[h]
    \centering
    \includegraphics[width=0.65\textwidth]{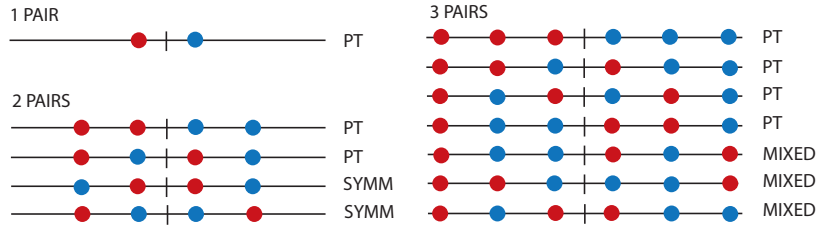}
    \caption{Illustration of the $\mathcal{PT}$-symmetric crystal lattice for one, two and three pairs of particles.}
    \label{fig1}
\end{figure}

In order to illustrate all three discussed cases: deterministic collection and random collections of types (A) and (B), by means of a single and simple example, we consider a situation in which $N$ gain and $N$ loss centers must precisely occupy the lattice points of a one-dimensional crystal structure. Different realizations are then created if the choices of lattice points occupied by gain and loss centres are changed. Note that for collections of passive particles such an example, being based only on switching but not on continuous motion, would lead to a degenerate situation in which the recombination would not result in creating qualitatively new realizations. Figure \ref{fig1} shows possible deterministic states formed by one, two and three pairs of gain/loss centres. Also, the congruent set of realizations is obtained by reflecting the states displayed with respect to the crystal's center (not shown). It is apparent that in order for such distributions to be $\mathcal{PT}$-symmetric, they have to obey condition (\ref{FPT}). Note also that starting from two-pair collections symmetric (passive) states, and starting from three-pair collections, mixed (partially $\mathcal{PT}$-symmetric) states are possible.

In order to construct a random, on average $\mathcal{PT}$-symmetric collection of type (A) it suffices to include any of the shown $\mathcal{PT}$ symmetric realizations with any probabilities. Also, to illustrate that a random collection constructed from $\mathcal{PT}$-symmetric realizations can be passive on average it suffices to use both realizations of the single-pair collection (shown and reflected) with equal probabilities. Further, to construct a random, on average $\mathcal{PT}$ symmetric collection of type (B) it suffices to cycle all four shown  realizations of the two-pair collection with equal probabilities (1/4). Even though the two last realizations are not $\mathcal{PT}$ symmetric the random collection is such on average (since the effects produced by two last realizations annihilate each other during the averaging process). Similarly, the pairs of unbalanced realizations being reflections of each other (not shown) produce effects that annihilate each other,  on average.

\begin{figure}[h]
    \centering
    \includegraphics[width=0.60\textwidth]{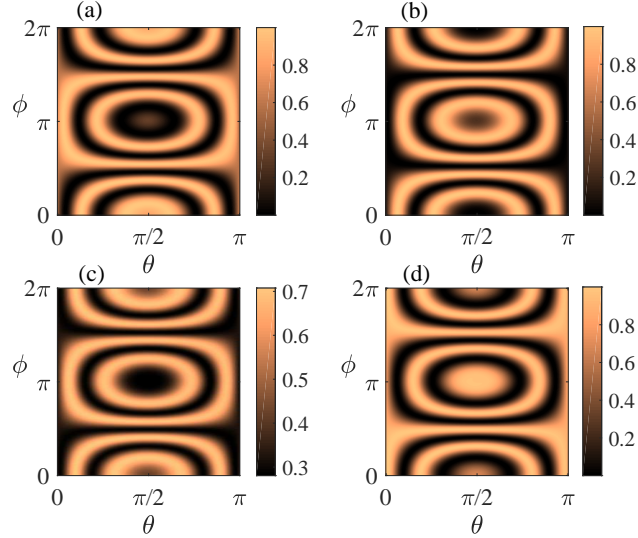}
    \caption{Spectral density $S^s(r\hat{s},\omega)$ normalized by factor $4S^i(\omega)B^2\sigma^3/r^2$, for a pair of particles with gain and loss described by \eqref{lastC}, versus polar and azimuthal scattering angles, $\phi$ and $\theta$, respectively. (a) $p = 0$, (b) $p = 1$, (c) $p = 1/2$ and (d) $\alpha = 0$ (passive case). Parameters used: $\hat{s}_0 = (1/\sqrt{3})(1,1,1)$, $\pmb{r}_0 = (6,0,0)$, $\alpha = 1$ (a-c), $k\sigma = 0.1$.}
    \label{fig2}
\end{figure}

More quantitatively, for a single-pair collection, a realization of the scattering potential can be constructed from the state set
\begin{equation}
F_{1,2}(\textbf{r},\omega)= f_M(\textbf{r}-\textbf{r}_0,\omega)e^{\pm i\alpha}+ f_M(\textbf{r}+\textbf{r}_0,\omega)e^{\mp i\alpha},
\end{equation}
with any real-valued, symmetric $f_M$, say Gaussian:  $f_M(\textbf{r})=B\exp\left[-r^2/2 \sigma^2 \right]$. Assume the Bernoulli distribution of realizations with probabilities of occurrence of mutually exclusive states $F_1$ and $F_2$ as $p$ and $1-p$, respectively, and let $J$/$J_1$ be the total/$F_1$ number of trials. Then: 
\begin{equation}
\begin{split}
A(\textbf{r},\omega) &=\lim\limits_{J\rightarrow \infty}\frac{1}{J} \sum\limits_{j=1}^{J}  F^j (\textbf{r},\omega)\\&
=\lim\limits_{J\rightarrow \infty}\frac{1}{J} \left[ J_1 F_1 (\textbf{r},\omega) + (J-J_1) F_2(\textbf{r},\omega) \right]\\&=p F_{1}(\textbf{r},\omega)+(1-p)F_{2}(\textbf{r},\omega).
\end{split}
\end{equation}
and 
\begin{equation}
\begin{split}
&C(\textbf{r}_1,\textbf{r}_2,\omega)=\lim\limits_{J\rightarrow \infty} \frac{1}{J} \sum\limits_{j=1}^J F^{j*}(\textbf{r}_1,\omega)F^j(\textbf{r}_2,\omega) \\&
=\lim\limits_{J\rightarrow \infty} \frac{1}{J} \Bigl[ J_1 F_1^* (\textbf{r}_1,\omega)F_1 (\textbf{r}_2,\omega) + (J-J_1) F_2^*(\textbf{r}_1,\omega)F_2(\textbf{r}_2,\omega)  \Bigr]\\&
= p F_1^* (\textbf{r}_1,\omega)F_1 (\textbf{r}_2,\omega) + (1-p) F_2^*(\textbf{r}_1,\omega)F_2(\textbf{r}_2,\omega).
\end{split}
\end{equation}
The spatial self-correlations $F^*_1F_1$ and $F^*_2F_2$ are perfect (if $F_1$ or $F_2$ occurs at $\textbf{r}_1$ it must occur at $\textbf{r}_2$ as well), hence they factorize, while the coefficients appear because of the frequency of occurrence in the summation. The cross-correlation terms are trivial since $F_1$ and $F_2$ are mutually exclusive for any $\textbf{r}$. For $p=1,0$ $\mathcal{PT}$-symmetric collections (with opposite sense) are obtained while for $p=1/2$ the collection degenerates to passive. Hence
\begin{equation}\label{lastC}
    \begin{split}
        \tilde{C}_F(\textbf{K}_1,\textbf{K}_2,\omega) &= 2\tilde{f}_M(\textbf{K}_1,\omega)\tilde{f}_{M}(\textbf{K}_2,\omega) \\
        &\times\Big\{ 2\cos^2\alpha \cos(\textbf{K}_1\cdot \textbf{r}_0)\cos(\textbf{K}_2\cdot \textbf{r}_0) \\
        &-2\sin^2\alpha \sin(\textbf{K}_1\cdot \textbf{r}_0)\sin(\textbf{K}_2\cdot \textbf{r}_0) \\
        &+ (1-2p)\sin 2\alpha \sin(\textbf{K}_1\cdot \textbf{r}_0)\cos(\textbf{K}_2\cdot \textbf{r}_0)\\
        &+ (2p-1)\sin 2\alpha \sin(\textbf{K}_2\cdot \textbf{r}_0)\cos(\textbf{K}_1\cdot \textbf{r}_0)\Big\},
    \end{split}
\end{equation}
from which one readily identifies the partial pair-structure factors and the corresponding coefficients [see Eqs. (\ref{PPP}) and (\ref{pp})]. Figure \ref{fig2} shows an example of light scattering from a system described by \eqref{lastC}. The strong dependence of the far-field spectral density on $p$ and $\alpha$ is evident from these plots.

\end{document}